\documentclass[aps, prl, twocolumn, amsmath, amssymb, a4paper, 10pt, superscriptaddress]{revtex4-2}

\usepackage{graphicx}% Include figure files
\usepackage[utf8]{inputenc}
\usepackage{bm}% bold math
\usepackage{xcolor}
\usepackage{physics}
\usepackage[colorlinks, citecolor=blue, linkcolor=blue]{hyperref}
\usepackage{siunitx}
\usepackage{xspace}
\usepackage{nicefrac}

\begin{document}

\newcommand{\mT}{\milli\tesla}
\newcommand{\mJmsq}{\milli\joule\per\square\meter}

\title{Thermal spin wave noise as a probe for the Dzyaloshinkii-Moriya interaction}

\author{Aurore Finco}
\email{aurore.finco@umontpellier.fr}
\author{Pawan Kumar}
\affiliation{Laboratoire Charles Coulomb, Université de Montpellier, CNRS, Montpellier, France}
\author{Van Tuong Pham}
\affiliation{Université Grenoble Alpes, CNRS, CEA, SPINTEC, 38054 Grenoble, France}
\affiliation{IMEC, 3001 Leuven, Belgium}
\author{Joseba Urrestarazu-Larrañaga}
\author{Rodrigo Guedas Garcia}
\affiliation{Université Grenoble Alpes, CNRS, CEA, SPINTEC, 38054 Grenoble, France}
\author{Maxime Rollo}
\affiliation{Laboratoire Charles Coulomb, Université de Montpellier, CNRS, Montpellier, France}
\author{Olivier Boulle}
\affiliation{Université Grenoble Alpes, CNRS, CEA, SPINTEC, 38054 Grenoble, France}
\author{Joo-Von Kim}
\affiliation{Centre de Nanosciences et de Nanotechnologies, CNRS, Université Paris-Saclay, 91120 Palaiseau, France}
\author{Vincent Jacques}
\affiliation{Laboratoire Charles Coulomb, Université de Montpellier, CNRS, Montpellier, France}

\begin{abstract}
  Interfacial Dzyaloshinkii-Moriya interaction (DMI) is a key ingredient in the stabilization of chiral magnetic states in thin films. Its sign and strength often determine crucial properties of magnetic objects, like their topology or how they can be manipulated with currents. A few experimental techniques are currently available to measure DMI quantitatively, based on the study of domain walls, spin waves, or spin-orbit torques. In this work, we propose a qualitative variant of spin wave methods. We rely on magnetic noise from confined thermal spin waves in domain walls and skyrmions in perpendicularly magnetized thin films, which we probe with scanning NV center relaxometry. We show both numerically and experimentally that the sign of the DMI can be inferred from the amplitude of the detected noise, which is affected by the non-reciprocity in the spin wave dispersion. Furthermore, we also demonstrate that the noise distribution around the contour of magnetic skyrmions reveals their Néel/Bloch nature, giving therefore also insight into the strength of DMI involved in their stabilization. 
\end{abstract}
\date{\today}

\maketitle

% intro
Magnetic noise provides valuable information about systems in which incoherent or thermal magnetic excitations, as well as fluctuating spins, play a crucial role~\cite{sinitsynTheorySpinNoise2016}. For example, noise spectroscopy can provide access to electron and hole spin dynamics in semiconductors~\cite{oestreichSpinNoiseSpectroscopy2005, crookerSpinNoiseElectrons2010}, properties of nanoparticles~\cite{jonssonDynamicStudyDipoledipole1998, stipeMagneticDissipationFluctuations2001a, tetienneSpinRelaxometrySingle2013, schmid-lorchRelaxometryDephasingImaging2015}, information on magnetic materials close to a phase transition~\cite{balkCriticalBehaviorZerofield2014}, the detection of thermal spin waves~\cite{vandersarNanometrescaleProbingSpin2015, duControlLocalMeasurement2017}, and spin transport phenomena~\cite{wangNoninvasiveMeasurementsSpin2022}. Quantum sensors like nitrogen-vacancy (NV) centers in diamond are especially useful to investigate these noise-related phenomena owing to their high sensitivity to their environment, whose fluctuations modify their spin relaxation properties~\cite{degenQuantumSensing2017}. Integrating an NV sensor into an atomic force microscope combines high sensitivity noise sensing with nanoscale spatial resolution~\cite{rondinMagnetometryNitrogenvacancyDefects2014}. This approach, referred to as scanning relaxometry, allows the realization of electrical conductivity maps~\cite{ariyaratneNanoscaleElectricalConductivity2018} or the investigation of magnetic textures, in particular, in low-moment materials like antiferromagnets~\cite{flebusProposalDynamicImaging2018}.

In this paper, we demonstrate how scanning relaxometry using NV centers can be harnessed to characterize the \emph{handedness} of chiral spin textures, which in thin ferromagnetic films are governed by the sign and strength of the Dzyaloshinskii-Moriya interaction (DMI)~\cite{dzyaloshinskii1964theory, moriyaAnisotropicSuperexchangeInteraction1960}. It is known that thermal spin waves within textures like domain walls lie within the GHz range, which affects the longitudinal spin relaxation time $T_1$ of the NV center~\cite{fincoImagingNoncollinearAntiferromagnetic2021}. The core idea of this work is that the nonreciprocal propagation of these spin waves, resulting from the DMI, also gives rise to different noise profiles above and below the film, from which information about the handedness of the spin textures can be deduced.

% main text

\begin{figure*}%[h]
    \centering
    \includegraphics[width=\linewidth]{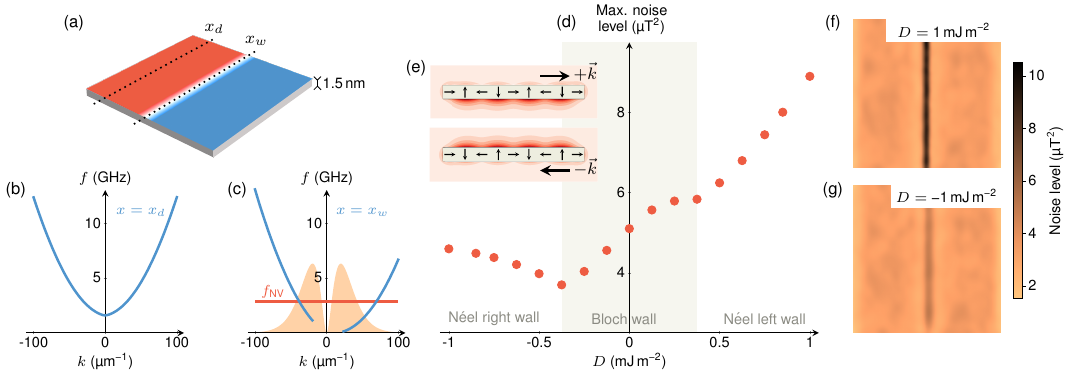}
    \caption{(a) Sketch of the considered out-of-plane magnetized Co layer. (b)-(c) Calculated spin wave dispersion in the Co layer with $D=\SI{1.0}{\mJmsq}$, in (b) the domains (at $x=x_d$) and in (c) the domain walls (at $x=x_w$). The orange shading in (c) represents the $k$-filtering function at a distance of \SI{50}{\nm}. No data is shown for small $k$ in the domain wall because a perfectly straight wall is unstable under these conditions. (d) Calculated maximal noise level at \SI{2.87}{\GHz} and at \SI{50}{\nm} from the film surface, when varying the DMI constant $D$. (e) Sketch of the stray field produced by Halbach arrays. (f)-(g) Calculated noise map \SI{50}{\nm} above the film, for $D = \pm \SI{1.0}{\mJmsq}$.}
    \label{fig:theory}
  \end{figure*}

We first illustrate the concept by considering an ultrathin ferromagnetic film in contact with a heavy-metal underlayer, which induces an interfacial DMI~\cite{fertRoleAnisotropicExchange1980, crepieuxDzyaloshinskyMoriyaInteractions1998}. This form of the interaction favors spin spirals propagating in the film plane~\cite{bodeChiralMagneticOrder2007}, which translates into an asymmetric dispersion relation for spin waves in the Damon-Eshbach geometry for in-plane magnetized films~\cite{udvardiChiralAsymmetrySpinWave2009, cortes-ortunoInfluenceDzyaloshinskiiMoriya2013, moonSpinWavePropagation2013}, but does not affect perpendicularly-magnetized systems in which propagation remains isotropic. This effect has been probed in Brillouin light scattering experiments in which the asymmetry in the spin wave dispersion relation~\cite{kuepferlingMeasuringInterfacialDzyaloshinskiiMoriya2023} can be measured to quantify the interfacial DMI. For a perpendicularly-magnetized film in a multi-domain state [Fig.~\ref{fig:theory}(a)], both reciprocal and nonreciprocal propagation are possible depending on where the propagation takes place. Within the domains, along the line indicated by $x_d$ in Fig~\ref{fig:theory}(a), the dispersion is reciprocal as shown in Fig.~\ref{fig:theory}(b). Along a \ang{180} Néel domain wall at the position $x_w$, on the other hand, the dispersion relation is strongly nonreciprocal with a narrow frequency gap as shown in Fig.~\ref{fig:theory}(c), where both the DMI and dipolar interactions contribute to the asymmetry~\cite{garcia-sanchezNarrowMagnonicWaveguides2015, henryUnidirectionalSpinWave2019}. The degree of asymmetry increases with $D$, as shown in Fig.~S2 of the Supplemental Material~\cite{suppl}.

Our objective is then to probe the magnetic noise produced by thermal spin waves inside such a domain wall using a single NV center placed above the magnetic film. Previous studies on synthetic antiferromagnets~\cite{fincoImagingNoncollinearAntiferromagnetic2021} have shown how the resulting fluctuating magnetic fields shorten the longitudinal spin relaxation time $T_1$, leading to a reduction of the photoluminescence (PL) signal emitted by the NV center~\cite{rolloQuantitativeStudyResponse2021}. This sensing method effectively filters the probed noise twice: it is frequency-filtered as only the frequency components around the magnetic resonance frequencies of the NV center [$f_\text{NV} \sim \SI{2.87}{\GHz}$ at zero field, depicted by the red line in Fig.~\ref{fig:theory}(c)], can enhance its relaxation, and it is wavevector-filtered as the NV sensitivity to fluctuations is broadly peaked around $\nicefrac{1}{d_\text{NV}}$, where $d_\text{NV}$ is the distance between the NV center and the magnetic film. This filter function $ke^{-2kd_\text{NV}}(1-e^{-2kt})$~\cite{duControlLocalMeasurement2017, vandersarNanometrescaleProbingSpin2015} is represented by the orange shading in Fig.~\ref{fig:theory}(c). This double filtering, combined with the asymmetric dispersion relation, means that the NV center is mostly sensitive to noise arising from modes with a given wavevector direction determined by the sign of $D$.

Note that no significant stray fields are expected from propagating spin waves in out-of-plane uniformly magnetized thin films; most reports to date in the literature~\cite{vandersarNanometrescaleProbingSpin2015, duControlLocalMeasurement2017, mccullianBroadbandMultimagnonRelaxometry2020, lee-wongNanoscaleDetectionMagnon2020, bertelliMagneticResonanceImaging2020, zhouMagnonScatteringPlatform2021, bertelliImagingSpinWaveDamping2021, simonFilteringImagingFrequencyDegenerate2022, koernerFrequencyMultiplicationCollective2022}, except Ref.~\onlinecite{fincoImagingNoncollinearAntiferromagnetic2021}, concern in-plane magnetized materials. In a perpendicularly magnetized film, the dynamic magnetization components of the propagating spin waves lie in the film plane, with $\vec{k}$ orthogonal to the static magnetization. These fluctuations do not produce surface or volume magnetic charges, so no fluctuating stray fields are expected.

Figure~\ref{fig:theory}(d) displays the expected noise level detectable by an NV center, positioned above a \ang{180}-domain wall at $d_\text{NV} = \SI{50}{\nm}$ from the film surface, as a function of $D$. These levels were computed following the method described in Refs.~\onlinecite{fincoImagingNoncollinearAntiferromagnetic2021, suppl}. Notably, we observe a significant difference in noise levels for left- and right-handed Néel walls. The conjunction of two phenomena gives rise to this effect. First, spin waves with opposite wave vectors about a static in-plane magnetization produce magnetic stray fields mainly above or below the film~\cite{devolderPropagatingspinwaveSpectroscopyUsing2023}. This can be understood by drawing an analogy between the magnetization fluctuations of the spin wave with the magnetic order within a Halbach array [Fig.~\ref{fig:theory}(e)]. The stray field generated by the arrangement of the in-plane magnetization compensates the contribution from the out-of-plane component on one side of the film but reinforces it on the other side, resulting in a one-sided flux magnetization configuration~\cite{mallinsonOnesidedFluxesMagnetic1973}. Reversing the handedness of the Halbach array results in the dominant flux being generated on the opposite film surface. Second, the DMI-induced nonreciprocity, $\omega(k) \neq \omega(-k)$, means that thermal populations of modes with opposite wave vector are unequal, i.e., $n(k) \neq n(-k)$ with $n(k) \simeq k_B T/\hbar \omega(k)$ in the long wavelength limit. This population imbalance grows as the dispersion becomes more asymmetric with the strength of $D$, which is reflected in the variation of the noise level. The noise maps in Figs.~\ref{fig:theory}(f) and (g) show the expected signal above Néel left and right domain walls, with $D= \pm \SI{1.0}{\mJmsq}$, at a distance of \SI{50}{\nm} from the film. A stronger noise contrast is expected for the Néel left wall.

We now turn to the experimental investigation of this effect using scanning NV center microscopy. We will measure static stray field maps by monitoring the Zeeman shift of the NV center magnetic resonance, and probe spin wave noise through the acceleration of the NV center relaxation and the subsequent PL decrease. Instead of the ferromagnetic single layer as discussed above, we study synthetic antiferromagnets (SAF) comprising two Co layers coupled antiferromagnetically through a Ru/Pt spacer~\cite{legrandRoomtemperatureStabilizationAntiferromagnetic2020, phamFastCurrentinducedSkyrmion2024}. This choice is motivated by the low static stray fields generated by the antiparallel magnetizations, i.e., $<\SI{1}{\mT}$ measured at the NV center position, in contrast to \SI{15}{\mT} at \SI{60}{\nm} above a domain wall in a single \SI{1.5}{\nm}-thick Co layer. Moreover, the stray field distribution from a ferromagnet possesses a strong off-axis component with respect to the NV axis, which corresponds to the direction joining the nitrogen atom and the vacancy. This off-axis field leads to a mixing of the NV center spin states, and a drop of the emitted PL~\cite{tetienneMagneticfielddependentPhotodynamicsSingle2012}. This drop would further compound the PL decrease induced by the shortening of $T_1$ by magnetic noise. Using a SAF in our experiment therefore allows us to avoid this unwanted effect and to focus solely on the response of the NV center to thermal spin wave noise. Nevertheless, experiments on a single Co layer were also performed (Fig.~S1~\cite{suppl}) and agree with our theoretical predictions. All our measurements are achieved under the application of a few \si{\mT} bias field used to separate the magnetic resonances of the NV center during quantitative measurements, but not affecting the magnetic texture or the noise detection, as the noise has a broad frequency spectrum.

\begin{figure}%[h]
    \centering
    \includegraphics[width=0.95\linewidth]{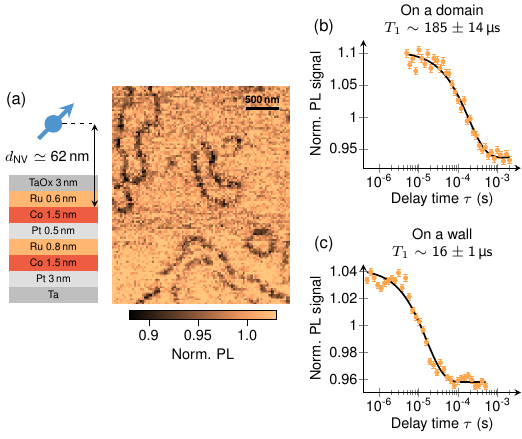}
    \caption{(a) Noise map of the SAF sketched on the left, measured with scanning NV center microscopy and revealing the presence of domain walls. (b)-(c) Relaxation time curves measured above a domain (b) or above a wall (c), showing a strong decrease of $T_1$ above the wall. Black lines are fits to an exponential decay.}
    \label{fig:DW_Left}
  \end{figure}

We perform the first experiment on a SAF grown on Si/SiO\textsubscript{2}, with the composition indicated in Fig.~\ref{fig:DW_Left}(a). We measure the PL emitted by the NV center as the tip is scanned over the sample surface at a distance $d_\text{NV} = \SI{62 \pm 5}{\nm}$ from the top Co layer (details of the tip calibration are given in Ref.~\onlinecite{suppl}). As in Ref.~\onlinecite{fincoImagingNoncollinearAntiferromagnetic2021} we observe a drop of PL above the domain walls, owing to the enhancement of the NV spin relaxation by the magnetic noise associated with spin waves localized within the wall. This enhancement of the relaxation rate is shown in Fig.~\ref{fig:DW_Left}(b) and (c), which compare measurements of $T_1$ taken above a domain and a wall, respectively (see ref.~\cite{suppl} for the measurement protocol) . In this sample, the wall type is expected to be left-handed Néel as the DMI arises from the Pt/Co interface, and was measured to be $D=\SI{0.62 \pm 0.24}{\mJmsq}$~\cite{phamFastCurrentinducedSkyrmion2024}. We are, therefore, in the favorable case for detection, with a strong noise contribution above the film.

\begin{figure}%[h]
    \centering
    \includegraphics[width=0.95\linewidth]{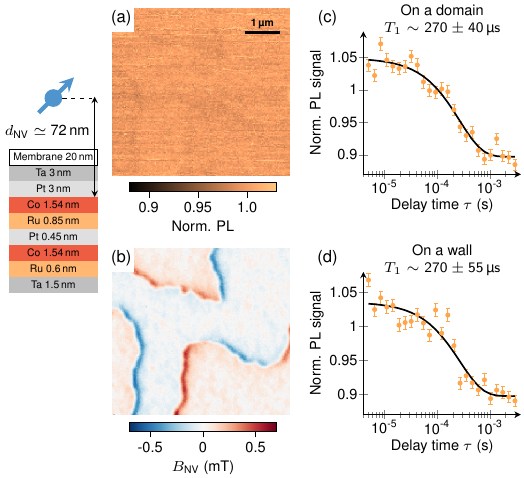}
    \caption{(a) Thermal noise map measurement on a reversed stack grown on a membrane. (b) Stray field map measured simultaneously with (b), showing the position of the domain walls. (c)-(d) Relaxation time curves measured above a domain (b) or above a domain wall (c). Black lines are fits to an exponential decay.}
    \label{fig:DW_Right}
  \end{figure}

Figure~\ref{fig:DW_Right} presents a similar experiment performed on an inverted stack, such that the sign of $D$ is reversed. We first attempted growing the multilayer stack in reverse order on Si/SiO\textsubscript{2}, but this resulted in weak perpendicular anisotropy in the Co layers such that it was not possible to stabilize perpendicularly-magnetized domains. Instead, we substituted the Si/SiO\textsubscript{2} substrate for an about \SI{20}{\nm}-thick SiN membrane, keeping the same growth order, and performed the scanning NV microscopy on the back of the sample through the membrane. To compensate for the thickness of the membrane, we use another diamond probe with an NV center closer to the surface, resulting in a net distance $d_\text{NV} = \SI{72 \pm 5}{\nm}$. The noise map obtained and shown in Fig.~\ref{fig:DW_Right}(a) is blank, in contrast to the clear 10\% decrease of PL above the walls in Fig.~\ref{fig:DW_Left}(a). Fig.~\ref{fig:DW_Right}(b) displays the static magnetic stray field measured simultaneously with the noise map in panel (a) and reveals the presence of domain walls in this area.

In order to investigate further this sample, we again measured $T_1$ above a uniformly-magnetized region and a domain wall (Figs.~\ref{fig:DW_Right} (c)-(d)). No significant difference is seen between these two experiments, which suggests that the detectable noise level is smaller for right-handed than left-handed Néel walls as predicted by our model. Our measurements performed on a single Co layer, with an identical diamond probe and identical $d_\text{NV}$ for both chiralities, also support this finding (Fig.~S1~\cite{suppl}), albeit with lower-quality measurements related to the stronger stray fields and difficulties encountered in scanning over the membrane. We have thus demonstrated that the noise level above Néel wall is indeed correlated to their handedness in samples hosting a significant interfacial DMI.

\begin{figure}%[h]
    \centering
    \includegraphics[width=\linewidth]{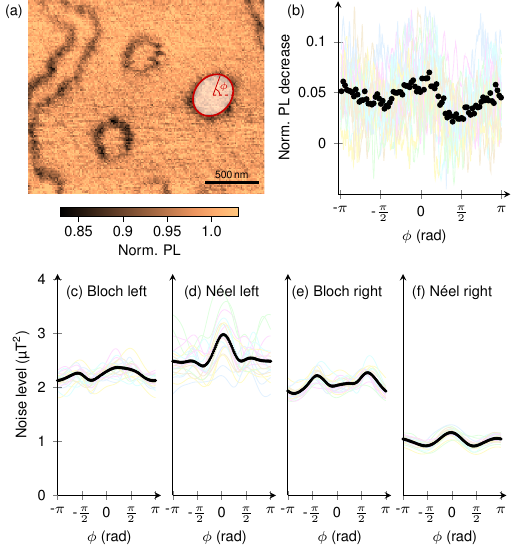}
    \caption{(a) Noise map showing three skyrmions. The PL signal is extracted along an elliptical contour of each skyrmion and plotted as a function of the angular position $\phi$. (b) Extracted PL decrease along the contour of 19 skyrmions, black dots show the average value, with stronger noise around $\phi=0$. (c)-(f) Expected noise distribution along the skyrmion contour for different skyrmion types. Thin colored lines are different disorder configurations, and black dots the mean value. In both the experiment and the calculations, the NV axis in-plane projection lies along $\phi=0$.}
    \label{fig:skyrmions}
  \end{figure}

Going a step further, we also studied skyrmions in SAF multilayers. Because DMI is critical for their existence, we anticipate that skyrmion noise maps can also provide signatures of the sign and strength of $D$. With the SAF sample shown in Fig.~\ref{fig:DW_Left}, skyrmions were nucleated by first applying an out-of-plane magnetic field of \SI{220}{\mT} and subsequently lowered to \SI{175}{\mT}~\cite{phamFastCurrentinducedSkyrmion2024}, after which the field is further reduced to a few \si{\mT}. These skyrmions remain pinned at low fields and possess a diameter of about \SI{200}{\nm}, which appear as dark rings in the noise maps in Figs.~\ref{fig:DW_Left}(a) and \ref{fig:skyrmions}(a). Interestingly, these dark rings do not exhibit a uniform PL intensity along their contours.

To better understand whether this non-uniformity is intrinsic to the internal spin texture or results only from disorder-related pinning, we compare the PL extracted along the contour of 19 skyrmions using an elliptical fitting function, as depicted in Fig.~\ref{fig:skyrmions}(a). We plot this data as a function of the angular position $\phi$ along the contour, with $\phi = 0$ corresponding to the horizontal direction on the images. Importantly, this horizontal direction corresponds to the in-plane projection of the NV axis, as determined by prior probe calibration.
The resulting plots are shown in Fig.~\ref{fig:skyrmions}(b) as thin colored lines. While each curve exhibits large fluctuations, as a consequence of both the PL noise and the spatial disorder that distorts the shape of the skyrmion, averaging over the ensemble (black dots) reveals a clear trend where the PL attains a minimum around $\phi = 0$.

We performed micromagnetics simulations of the magnetic noise in the SAF skyrmions following the method in Fig.~\ref{fig:theory}. We assumed $d_\text{NV} = 50$ nm and considered 20 different realizations of magnetic disorder, comprising a 1\% fluctuation in the perpendicular anisotropy, such that we obtain an ensemble of pinned skyrmions with diameters of about \SI{200}{\nm} but with varying shapes due to the disorder. This procedure was repeated for Bloch and Néel skyrmions with opposite chiralities. The calculated noise distributions are shown in Figs.~\ref{fig:skyrmions}(c)-(f), with individual contours shown as thin colored lines and the average as black dots. Note that only the spatial component of the noise perpendicular to the NV axis contributes to the acceleration of the spin relaxation~\cite{degenQuantumSensing2017, rolloQuantitativeStudyResponse2021} and therefore to the noise contrast. This is the only element in the experiment breaking the circular symmetry, allowing the observation of specific noise patterns depending on the internal structure of the skyrmions. In agreement with the experimental data, we observe that a stronger noise level is expected around $\phi=0$ (again the direction of the NV axis projection), resulting in the maximum observed in the PL decrease in Fig.~\ref{fig:skyrmions}(b). We also notice that a significantly lower noise level is expected for Néel right than for Néel left skyrmions, similar to what occurs in domain walls, since it is again the in-plane magnetized region of the skyrmion boundary that contributes to the noise. The predicted variation for Bloch skyrmions is qualitatively different from the measured signal, consistent with the Néel configuration of the skyrmions. 

%%
% Conclusion
%%
We have shown that a combination of physical mechanisms allows insights into the sign and magnitude of DMI in thin films to be gleaned using scanning NV center microscopy. The asymmetric spin wave dispersion driven by the DMI, along with the wavevector-dependent stray fields in thin films, produces magnetic noise associated with spin textures like domain walls and skyrmions that encodes the sign and strength of the DMI. This approach has the potential to be utilized, at least for samples that produce moderate stray fields such as ferrimagnets and antiferromagnets, to determine the Bloch/Néel nature of skyrmions without the need for precise quantitative stray field measurements. Our findings unveil new possibilities for leveraging NV center microscopy not only for efficient characterization of magnetic materials but also for magnonics, as it provides access to the properties of spin waves confined within nanoscale magnetic textures, which are challenging to investigate using conventional tabletop experimental methods.

\vspace*{3mm}
\begin{acknowledgements}
  The data that support this work are available in Zenodo with the DOI 10.5281/zenodo.14808937.
  
   \vspace*{3mm}
  We acknowledge support from the European Union’s Horizon 2020 research and innovation programme under grant agreements No. 964931 (TSAR) and No. 866267 (EXAFONIS). For the purpose of Open Access, a CC-BY public copyright licence
has been applied by the authors to the present document and will be applied to all subsequent versions up to the Author Accepted Manuscript arising from this submission. 
\end{acknowledgements}

%\bibliography{chirality_SW_noise.bib}
%\bibliographystyle{apsrev4-2}


\end{document}